# Description of electromagnetic fields in inhomogeneous accelerating sections. III Beam loading

## M.I. Ayzatsky[1]

National Science Center "Kharkiv Institute of Physics and Technology" (NSC KIPT), 61108, Kharkiv, Ukraine

A self-consistent semi-analytical theory of beam loading in inhomogeneous accelerating structures based on the generalized theory of coupled modes is proposed. A single-mode approximation was used when the fields are represented as a sum of two components, one of which is associated with the right travelling eigen wave, and the second with the left. However, this second component is not always a left travelling. When a field is excited by an electron beam it can have complex spatial distribution. The results of calculation of the distribution of electric fields excited by a relativistic electron beam are presented.

## 1. Introduction

There were attempts to construct a self-consistent semi-analytical theory of excitation of a nonuniform accelerating structures based on Maxwell's equations [1,2,3] similar to simple differential equations of a homogeneous waveguide [4,5,6], but they turned out to be not entirely consistent. Researchers mainly limited themselves to approximate equations valid for smooth nonuniform waveguides (see, for example, [7,8,9,10,11,12,13]) or a coupled cavity model (see, for example, [14,15,16,17])

Recently it was proposed to use a modified uniform basis for description of non-periodic structured waveguides and a new generalization of the theory of coupled modes was constructed [18,19,20,21]. One of the positive features of proposed coupled modes theory is the simple and clear procedure for taking into account the beam loading and the easy transition to the case of a homogeneous waveguide [4,5]. It was shown that an infinitive system of equations can be reduced and the single mode representation describes the electromagnetic fields with small errors in the regular part of the accelerating section [20,21]. Within the single-mode approximation, the fields are represented as the sum of two components, one of which is associated with the right-travelling eigenwave and the other with the left-travelling eigenwave.

It was shown that the second component in the absence of an electron beam can be either a right- travelling or a left- travelling wave [22]. However, at the operating frequency (tuning frequency) the component associated with the left-travelling eigenwave represents a right-travelling wave (increasing dependence of the phase on $z$) with the same phase shift as the design one. This fact is very important for understanding the accuracy of the phase tuning of the inhomogeneous accelerator sections. Indeed, if it represented a left-travelling wave (reflected one) there would be a standing wave component which gives additional errors in a phase shifts.

In this paper we propose a theory of beam loading in inhomogeneous accelerating structures based on the generalized theory of coupled modes. The results of calculation of the distribution of electric fields excited by a relativistic electron beam are presented.

## 2. Main equations

We will assume that the dependence of all quantities on time is $\exp(-i\omega t)$.

Electromagnetic fields in a non-periodic structured waveguide with the ideal metal walls can be represented in the form of such series ([18])

$$\vec{H}(\vec{r}) = \sum_{s=-\infty}^{s=\infty} C_s(z)\vec{H}_s^{(e,z)}(\vec{r}),$$

$$\vec{E}(\vec{r}) = \sum_{s=-\infty}^{s=\infty} C_s(z)\vec{E}_s^{(e,z)}(\vec{r}) + \frac{\vec{j}}{i\omega\varepsilon_0\varepsilon}, \tag{1}$$

[1] E-mail: mykola.aizatsky@gmail.com, aizatsky@kipt.kharkov.ua

where $\vec{E}_s^{(e,z)}(\vec{r})$, $\vec{H}_s^{(e,z)}(\vec{r})$ are modified eigen vector functions obtained by generalizing the eigen $\vec{E}_s^{(e)}$, $\vec{H}_s^{(e)}$ vectors of a homogeneous waveguide by special continuation of the geometric parameters ([18,19]). The eigen waves of a homogeneous waveguide we present as $\left(\vec{E}_s, \vec{H}_s\right) = \left(\vec{E}_s^{(e)} \vec{H}_s^{(e)}\right)\exp\left(\gamma_s z\right)$, where $\left(\vec{E}_s^{(e)} \vec{H}_s^{(e)}\right)$ are the periodic functions of the z-coordinate. Under such choice of the basis functions, the coefficients $C_s(z)$ include an exponential dependence on the z-coordinate.

Electromagnetic fields (1) must obey the Maxwell's equations. To satisfy this condition $C_s(z)$ must be solutions of such a coupled system of differential equations [18,19,20,21]

$$\frac{dC_s}{dz} - \gamma_s^{(e,z)} C_s + \frac{1}{2N_s^{(e,z)}}\frac{dN_s^{(e,z)}}{dz} C_s + \sum_{s'=-\infty}^{\infty} C_{s'} U_{s',s}^{(z)} = \frac{1}{N_s^{(e,z)}}\int_{S_\perp^{(z)}} \vec{j}\vec{E}_{-s}^{(e,z)} dS, \qquad (2)$$

where

$$U_{s',s} = \frac{1}{2N_s^{(z)}}\sum_i \frac{dg_i^{(z)}}{dz}\int_{S_i^{(z)}(z)}\left\{\left[\frac{\partial \vec{E}_{-s}^{(e,z)}}{\partial g_i^{(z)}}\vec{H}_{s'}^{(e,z)}\right] - \left[\vec{E}_{-s}^{(e,z)}\frac{\partial \vec{H}_{s'}^{(e,z)}}{\partial g_i^{(z)}}\right] + \left[\frac{\partial \vec{E}_{s'}^{(e,z)}}{\partial g_i^{(z)}}\vec{H}_{-s}^{(e,z)}\right] - \left[\vec{E}_{s'}^{(e,z)}\frac{\partial \vec{H}_{-s}^{(e,z)}}{\partial g_i^{(z)}}\right]\right\}\vec{e}_z dS, \quad (3)$$

$g_i^{(z)}(z)$ -generalized geometrical parameters, $\gamma_s^{(e,z)}(z)$ - generalized wavenumber and

$$N_s^{(z)} = \int_{S_\perp^{(z)}(z)}\left\{\left[\vec{E}_s^{(e,z)}\vec{H}_{-s}^{(e,z)}\right] - \left[\vec{E}_{-s}^{(e,z)}\vec{H}_s^{(e,z)}\right]\right\}\vec{e}_z dS = \begin{cases} N_s^{(z)}, & s > 0, \\ -N_s^{(z)}, & s < 0. \end{cases} \qquad (4)$$

We will consider axisymmetric TH (E) electromagnetic fields. In this case $\vec{E}_s^{(e,z)} = \vec{e}_r E_{r,s}^{(e,z)} + \vec{e}_z E_{z,s}^{(e,z)}$ and $\vec{H}_s^{(e,z)} = \vec{e}_\varphi H_{\varphi,s}^{(e,z)}$.

In the single mode approach ($s = 1$) [19,20], when the representation (1) transforms into

$$\vec{E}_1(\vec{r}) = \vec{E}_1^+(\vec{r}) + \vec{E}_1^-(\vec{r}) = C_1(z)\vec{E}_1^{(e,z)}(\vec{r}) + C_{-1}(z)\vec{E}_{-1}^{(e,z)}(\vec{r}), \qquad (5)$$

the coupled system (2) is written as

$$\frac{dC_1}{dz} - \gamma_1^{(e,z)} C_1 + \frac{1}{2N_s^{(e,z)}}\frac{dN_s^{(e,z)}}{dz} C_1 + C_1 U_{1,1} + C_{-1} U_{-1,1} = \frac{1}{N_1^{(e,z)}}\int_{S_\perp^{(z)}} \vec{j}\vec{E}_{-1}^{(e,z)} dS$$

$$\frac{dC_{-1}}{dz} + \gamma_1^{(e,z)} C_{-1} + \frac{1}{2N_s^{(e,z)}}\frac{dN_s^{(e,z)}}{dz} C_2 + C_{-1} U_{-1,-1} + C_1 U_{1,-1} = -\frac{1}{N_1^{(e,z)}}\int_{S_\perp^{(z)}} \vec{j}\vec{E}_1^{(e,z)} dS \qquad (6)$$

We will consider $\vec{E}_k^{(e,z)}$ as dimensionless vector, the vector $\vec{H}_k^{(e,z)}$ has a dimension $[\vec{H}_k^{(e,z)}] = \Omega^{-1}$.

Taking into account that $U_{-1,-1} = -U_{1,1}$ and introducing new functions $\tilde{C}_{\pm 1}$,

$$C_{\pm 1} = E_D \sqrt{\frac{N_1^{(e,z)}(0)}{N_1^{(e,z)}(z)}}\Gamma^{(\pm)}(z)\tilde{C}_{\pm 1}, \qquad (7)$$

where

$$\Gamma^{(\pm)}(z) = \exp\left(\pm\int_0^z\left(\gamma_1^{(e,z)} - U_{1,1}\right)dz'\right), \qquad (8)$$

the system (6) takes the form

$$\frac{d\tilde{C}_1}{dz} = -\Gamma^{(-)2} U_{-1,1}\tilde{C}_{-1} + \frac{\Gamma^{(-)}}{E_D\sqrt{N_1^{(e,z)}(0)N_1^{(e,z)}(z)}}\int_{S_\perp^{(z)}} \vec{j}\vec{E}_{-1}^{(e,z)} dS,$$

$$\frac{d\tilde{C}_{-1}}{dz} = -\Gamma^{(+)2} U_{1,-1}\tilde{C}_1 - \frac{\Gamma^{(+)}}{E_D\sqrt{N_1^{(e,z)}(0)N_1^{(e,z)}(z)}}\int_{S_\perp^{(z)}} \vec{j}\vec{E}_1^{(e,z)} dS. \qquad (9)$$

$E_D$ is a coefficient which has the dimension $[E_D] = $ V/m. In the following we take $E_D = 1$ MV/m and the values of $C_{\pm 1}$ and $\vec{E}_1^\pm$ in all figures are given in MV/m.

The accelerating section under consideration, which is an analogue of the one developed at CERN [23] consists of $N_R$ =27 regular sells [20,21]. The couplers on both sides of the accelerating section were modeled by homogeneous structured waveguides whose dimensions coincide with the dimensions of the start and end cells. The electron beam interacts only with $N_R$ regular sells ($0 < z < L = N_R D$, $D$ =const - the period of the waveguide). In the left homogeneous waveguide, there may be a right traveling wave (wave from an external source), but in the right homogeneous waveguide the left traveling wave is absent.

For simplicity, we will assume that the electron beam at the entrance to the section ($z = 0$) is an infinite sequence of point bunches with charge $Q$, the entry time of which is equal to $t_{0,l} = lT_0$, where $l$ is the bunch number. Each bunch moves along the axis with a velocity $\vec{v} = \vec{e}_z c$. Such beam will excite fields with the frequencies $\omega = \frac{2\pi m}{T_0} = m\omega_0$. We will assume that the frequency $\omega_0$ lies within the first passband and $\omega_0 D / c = \theta_0 = 2\pi / 3$. In this case field harmonics with $m = \pm 1$ will be the largest and we will consider only them. Then we can write (see Appendix 1)

$$\vec{j} = \vec{e}_z j_z = \vec{e}_z \frac{Q}{T_0} \delta(x)\delta(y) \exp\left\{ i\frac{\omega_0 z}{c}m \right\} \tag{10}$$

and the system takes the form ($m = 1$)

$$\frac{d\tilde{C}_1}{dz} = -\Gamma^{(-)2} U_{-1,1}\tilde{C}_{-1} + \frac{QZ_0}{E_D} \frac{\Gamma^{(-)} E_{-1,z}^{(e,z)}(r=0,z)}{\sqrt{\tilde{N}_1^{(e,z)}(0)}\tilde{N}_1^{(e,z)}(z)} \exp\left( i\frac{\omega_0 z}{c} \right),$$

$$\frac{d\tilde{C}_{-1}}{dz} = -\Gamma^{(+)2} U_{1,-1}\tilde{C}_1 - \frac{QZ_0}{T_0 E_D} \frac{\Gamma^{(+)} E_{1,z}^{(e,z)}(r=0,z)}{\sqrt{\tilde{N}_1^{(e,z)}(0)}\tilde{N}_1^{(e,z)}(z)} \exp\left( i\frac{\omega_0 z}{c} \right), \tag{11}$$

where $\tilde{N}_1^{(e,z)}(z) = N_1^{(e,z)}(z)Z_0$, $Z_0 = \sqrt{\mu_0 / \varepsilon_0}$ .

To solve the system of first-order differential equations(11), it is necessary to formulate the boundary conditions. They are difficult to formulate for a real accelerator section, which usually has couplers (two for the traveling wave section and one for the standing wave section). As we model the couplers by homogeneous structured waveguides, we can use the open boundaries conditions. Boundaries conditions are formulated for the travelling wave section as follows:

$$\tilde{C}_1(z=0) = C_0, \tilde{C}_{-1}(z=L) = 0 . \tag{12}$$

A general method for numerical analysis of the problem of wave propagation and scattering in inhomogeneous plane-layered media is the finite difference method (see, for example, [24,25]). We used more accurate method based on the 4th order Runge-Kutta method [26,27].

The system (11) has two peculiarities. First peculiarity is the presence of an integral $\Gamma^{(\pm)}(z)$ with a variable upper limit. To calculate it, we used Simpson formula. Second, since the thickness of the disks ($d_{1,k}$) and the length of the resonators ($d_{2,k}$) are not constant and differ significantly[2] we cannot use the mesh with constant step. We chose the mesh $\{z_n\}$, $n = 1 \div N$, where the number of divisions of each segment (disk, resonator) $N_D$ is constant. We usually took $N_D$ =60. With such discretization, the step is not constant along the structure. To implement the Runge-Kutta method for each segment with a step $h_k$ it is necessary to divide the interval $[z_n, z_{n+1} = z_n + h_k]$ into four equal parts and, using the Simpson formular, calculate the values of the integral and right parts in two points: $z = z_n + (z_{n+1} - z_n)/2$ and $z = z_{n+1}$. The number of grid nodes used in the Runge-Kutta method will be equal $N = 2N_D N_R$ (the total number of grid nodes $2N$)

The resulting system of linear equations has the form (see Appendix 2)

---

[2] Iris thickness $d_{1,s}$ =1.67 ÷ 1.00 mm, structure period $D = d_{s,1} + d_{s,2}$ =8.3317 mm [23].

$$-\alpha_{1,2}(1)R + \tilde{C}_{1,2} = \alpha_{1,1}(1)C_0 + W_{1,1}$$

$$-\alpha_{2,2}(1)R + \tilde{C}_{2,2} = \alpha_{2,1}(1)C_0 + W_{2,1}$$

$$\left.\begin{array}{l} \alpha_{1,1}(n)\tilde{C}_{1,n} + \alpha_{1,2}(n)\tilde{C}_{2,n} - \tilde{C}_{1,n+1} = -W_{1,n} \\ \alpha_{2,1}(n)\tilde{C}_{1,n} + \alpha_{2,2}(n)\tilde{C}_{2,n} - \tilde{C}_{2,n+1} = -W_{2,n} \end{array}\right\} \quad n = 2,\ldots,N-2 \tag{13}$$

$$\alpha_{1,1}(N-1)\tilde{C}_{1,N-1} + \alpha_{1,2}(N-1)\tilde{C}_{2,N-1} - T = -W_{1,N-1}$$

$$\alpha_{2,1}(N-1)\tilde{C}_{1,N-1} + \alpha_{2,2}(N-1)\tilde{C}_{2,N-1} = -W_{2,N-1}$$

where $\tilde{C}_{1,n} = \tilde{C}_1(z_n)$, $\tilde{C}_{2,n} = \tilde{C}_2(z_n)$ and $\tilde{C}_{1,1} = C_0$, $\tilde{C}_{2,1} = R$, $\tilde{C}_{1,N} = T$, $\tilde{C}_{2,N} = 0$.

System of equations (13) is sparce, so to solve it we used a special numerical procedures LSLZG from the IMSL MATH LIBRARY.

### 3. Calculation results

In this paper the results of a study of the field distribution arising when point relativistic bunches with a charge $Q$ and constant velocity ($v_0 = c$) pass through 27 cells are given. System (13) is linear, so the fields from an external RF source and the beam are summed. The field distribution in an inhomogeneous structured waveguide excited by an external RF source was studied in [19,20]. Here we will consider only the beam fields ($C_0 = 0$).

#### 3.1 Beam loading in a homogeneous waveguide

Let us first consider the case of a homogeneous waveguide with geometric parameters that coincide with the parameters of the initial cell of a structure under consideration: iris thickness $d_{1,s} = 1.67$ mm, iris radii $b_{k,s} = 3.15$ mm. For homogeneous waveguide $U_{1,1} = U_{1,-1} = U_{-1,1} = 0$ and the equations of system (6) become independent and coincide with the known ones [4,5,6]. Moreover, $\tilde{N}_1^{(e,z)}(z) = const$, $\gamma_1^{(e,z)} = const$ (see Figure 1-Figure 3), $\Gamma^{(\pm)}(z) = \exp(\pm\gamma_1^{(e,z)}z)$, $\vec{E}_{\pm1}^{(e,z)}(r=0,z) = \vec{E}_{\pm1}^{(e,z)}(r=0,z)$ are periodic functions. In this case the amplitudes $C_1(z)$ and $C_{-1}(z)$, and hence the two field components $\vec{E}_1^+(\vec{r})$ and $\tilde{E}_1^-(\vec{r})$, are independent.

If we expand $E_{\pm1,z}^{(e,z)}(r=0,z)$ into a Fourie series

$$E_{\pm1,z}^{(e,z)}(r=0,z) = \sum_k G_{\pm1,k}^{(e)} \exp\left(i\frac{2\pi k}{D}z\right), \tag{14}$$

we can find, using the boundary conditions (12), the analytical expressions for $C_1(z)$ and $C_{-1}(z)$

$$C_1 = \frac{QZ_0 D}{E_D \tilde{N}_1^{(e)}} \exp\left(\gamma_1^{(e)}z\right) \sum_k G_{-1,k}^{(e)} \frac{\exp\left(i\theta_0 z - \gamma_1^{(e)}z + i\frac{2\pi k}{D}z\right) - 1}{i\theta_0 - \gamma_1^{(e)}D + i2\pi k}, \tag{15}$$

$$C_{-1} = \frac{QZ_0 D}{T_0 E_D \tilde{N}_1^{(e)}} \exp\left(-\gamma_1^{(s)}z\right) \sum_k G_{1,k}^{(e)} \frac{\exp\left(i\theta_0 L + \gamma_1^{(e)}L + i\frac{2\pi k}{D}L\right) - \exp\left(i\theta_0 z + \gamma_1^{(e)}z + i\frac{2\pi k}{D}z\right)}{i\theta_0 + \gamma_1^{(e)}D + i2\pi k}. \tag{16}$$

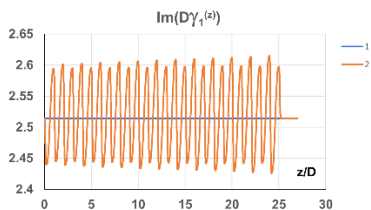

*Figure 1 Dependences on $z$ of the imaginary part of the local logitudinal wavenumber: 1- constant impedance case, 2 - nonuniform case.*

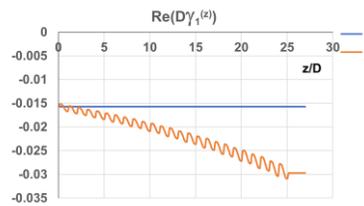

*Figure 2 Dependences on $z$ of the real part of the local logitudinal wavenumber: 1 - constant impedance case, 2 - nonuniform case.*

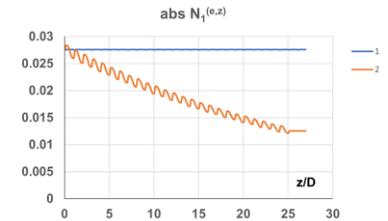

*Figure 3 Dependences on $z$ of the norm $\tilde{N}_1^{(e,z)}(z)$ : 1- constant impedance case, 2 - nonuniform case*

Numerical calculations show that in our case in the series (14) we can leave only two members: for $E_1^{(e)}(r=0,z)$ with $k=0$, $k=-1$ and for $E_{-1}^{(e)}(r=0,z)$ with $k=0$, $k=+1$[3]. Then we can rewrite (15) and (16) as

$$C_{\pm 1} = \frac{QZ_0 D}{E_D \tilde{N}_1^{(e)}} G_{\mp 1,0}^{(e)} \Lambda^{\pm}(z) = \frac{QZ_0 D}{E_D \tilde{N}_1^{(e)}} G_{\mp 1,0}^{(e)} \left( \Lambda_1^{\pm}(z) + \Lambda_2^{\pm}(z) \right), \tag{17}$$

where

$$\Lambda^{\pm}(z) = \Lambda_1^{\pm}(z) + \Lambda_2^{\pm}(z), \tag{18}$$

$$\Lambda_1^+(z) = \exp\left(\gamma_1^{(e)} z\right) \frac{\exp\left(i\theta_0 z / D - \gamma_1^{(e)} z\right) - 1}{i\theta_0 - \gamma_1^{(e)} D},$$

$$\Lambda_2^+(z) = \frac{G_{-1,1}^{(e)}}{G_{-1,0}^{(e)}} \exp\left(\gamma_1^{(e)} z\right) \frac{\exp\left(i\theta_0 z / D - \gamma_1^{(e)} z + i2\pi z / D\right) - 1}{i\theta_0 - \gamma_1^{(e)} D + i2\pi}, \tag{19}$$

$$\Lambda_1^-(z) = \exp\left(i\theta_0 L / D\right) \exp\left(-\gamma_1^{(s)}(z - L)\right) \frac{1 - \exp\left(i\theta_0(z-L) / D + \gamma_1^{(s)}(z-L)\right)}{i\theta_0 + \gamma_1^{(s)}},$$

$$\Lambda_2^-(z) = \exp\left(i\theta_0 L / D\right) \frac{G_{1,-1}}{G_{1,0}} \exp\left(-\gamma_1^{(s)}(z-L)\right) \frac{1 - \exp\left(i\theta_0(z-L) / D + \gamma_1^{(s)}(z-L) - i2\pi(z-L) / D\right)}{i\theta_0 + \gamma_1^{(s)} D - i2\pi}. \tag{20}$$

Results of calculations of the dependencies of the functions $\Lambda_1^+(z)$, $\Lambda_2^+(z)$, $\Lambda_1^-(z)$, $\Lambda_2^-(z)$, $\Lambda^-(z)$ on $z$ are presented in Figure 4-Figure 21.

Part of the field that describes by function $\Lambda_1^+(z)$ (see Figure 4-Figure 6) is well known in accelerator physics. It describes the wave with monotonically growth of amplitude (for small values of parameter $\left| z \operatorname{Re} \gamma_1^{(o)} \right|$ the growth is linear) and with a phase shift per cell $\theta = 2\pi / 3$.

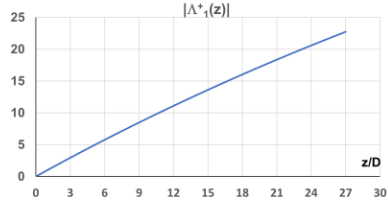

*Figure 4 Spatial distributions of the modulus of the function $\Lambda_1^+(z)$.*

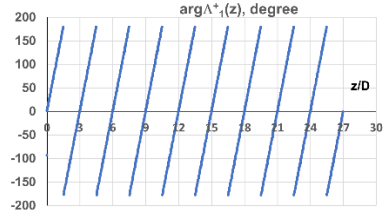

*Figure 5 Spatial distributions of the phase of the function $\Lambda_1^+(z)$.*

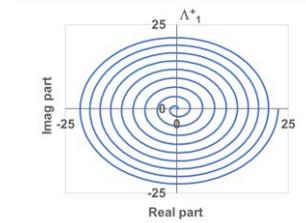

*Figure 6 Trajectory in the complex plane of the function $\Lambda_1^+(z)$.*

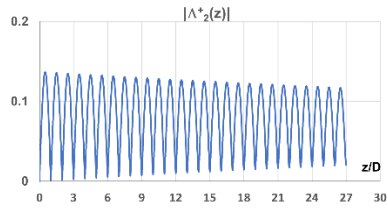

*Figure 7 Spatial distributions of the modulus of the function $\Lambda_2^+(z)$.*

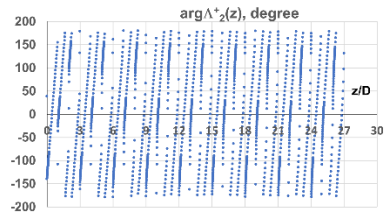

*Figure 8 Spatial distributions of the phase of the function $\Lambda_2^+(z)$.*

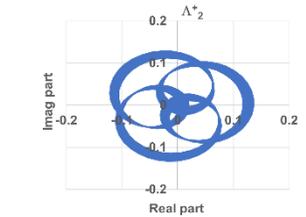

*Figure 9 Trajectory in the complex plane of the function $\Lambda_2^+(z)$.*

Part of the field that describes by function $\Lambda_2^+(z)$ (see Figure 7-Figure 9) has more complicated behavior. It determined by a combination of three wave components: beam component $\exp\left(i\theta_0 \frac{z}{D}\right)$, eigen wave component $\exp\left(\gamma_1^{(e)} z\right)$ and periodicity component $\exp\left(i\frac{2\pi}{D} z\right)$. When $\left| z \operatorname{Re} \gamma_1^{(e)} \right| << 1$ we have

$$\Lambda_2^+(z) \simeq \frac{G_{-1,1}^{(e)}}{\pi G_{-1,0}^{(e)}} \exp\left(i\frac{\theta_0 + \pi}{D}z\right)\sin\left(\frac{\pi}{D}z\right) = \frac{G_{-1,1}^{(e)}}{\pi G_{-1,0}^{(e)}}\exp\left(i\frac{5\pi}{3D}z\right)\sin\left(\frac{\pi}{D}z\right). \qquad (21)$$

---

[3] For example, $\left| G_{1,-2}^{(e)} \right| = $ 2.06E-2, $\left| G_{1,-1}^{(e)} \right| = 0.31$, $\left| G_{1,0}^{(e)} \right| = 0.72$, $\left| G_{1,1}^{(e)} \right| = $ 1.99E-2, $\left| G_{1,2}^{(e)} \right| = $ 2.73E-3.

Despite the presence of $\exp(i5\pi z/D)$ in the formula (21), $\Lambda_2^+(z)$ has a phase shift per cell equal to $2\pi/3$ (see Figure 10-Figure 12) which is determined by the change of sign of $\sin(\pi z/D)$ at the end of each cell. It follows from this that the part of the field described by the function $\Lambda_2^+(z)$ represent field that is a symbiosis of travelling and standing waves. Since the modulus of $\Lambda_2^+(z)$ is much smaller that the modulus $\Lambda_1^+(z)$, the function $\Lambda^+(z)$ differs from $\Lambda_1^+(z)$ only in a few initial sells.

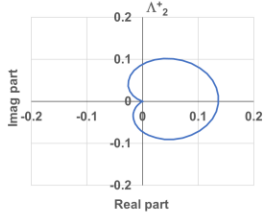

*Figure 10 Trajectory in the complex plane of the function $\Lambda_2^+(z)$, 15D<z<16D, $\left|\operatorname{Re}\gamma_1^{(e)}\right| = 0$.*

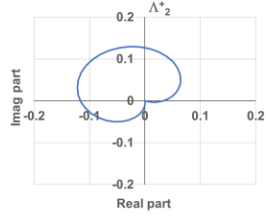

*Figure 11 Trajectory in the complex plane of the function $\Lambda_2^+(z)$, 16D<z<17D, $\left|\operatorname{Re}\gamma_1^{(e)}\right| = 0$.*

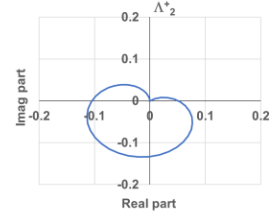

*Figure 12 Trajectory in the complex plane of the function $\Lambda_2^+(z)$, 17D<z<18D, $\left|\operatorname{Re}\gamma_1^{(e)}\right| = 0$.*

The part of the field described by the function $\Lambda_1^-(z)$ (see Figure 13-Figure 15) represents travelling-standing wave pattern with a phase shift per cell equal to $2\pi/3$, but with an amplitude several times greater than the amplitude of $\Lambda_2^+(z)$.

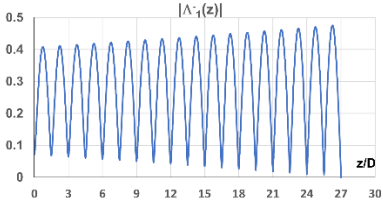

*Figure 13 Spatial distributions of the modulus of the function $\Lambda_1^-(z)$.*

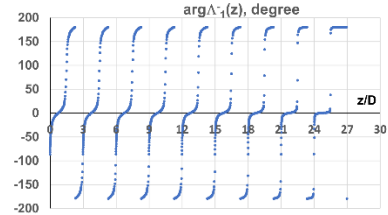

*Figure 14 Spatial distributions of the phase of the function $\Lambda_1^-(z)$.*

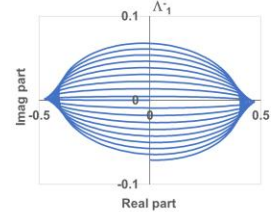

*Figure 15 Trajectory in the complex plane of the function $\Lambda_1^-(z)$.*

The part of the field described by the function $\Lambda_2^-(z)$ (see Figure 16-Figure 18) represents travelling-standing wave pattern with a phase shift per cell equal to $(-4\pi/3)$ and with the period of amplitude oscillations equals to the length of three cells (3D).

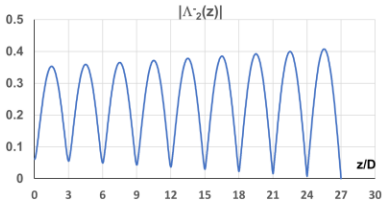

*Figure 16 Spatial distributions of the modulus of the function $\Lambda_2^-(z)$.*

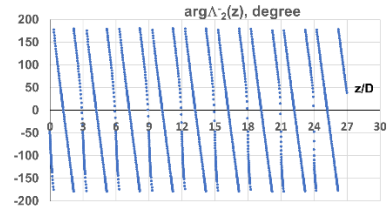

*Figure 17 Spatial distributions of the phase of the function $\Lambda_2^-(z)$.*

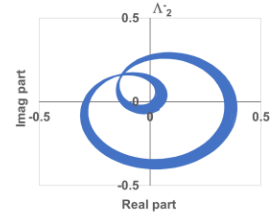

*Figure 18 Trajectory in the complex plane of the function $\Lambda_2^-(z)$.*

The total function $\Lambda^-(z) = \Lambda_1^-(z) + \Lambda_2^-(z)$ (see Figure 19-Figure 21) represents more complicated pattern than the travelling-standing one. Within every three cells the phase changes in an interval that is less than $2\pi$.

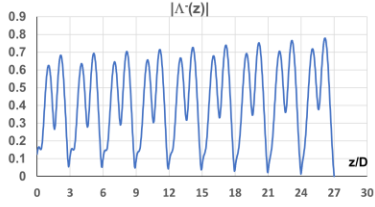

Figure 19 Spatial distributions of the modulus of the function $\Lambda^-(z)$.

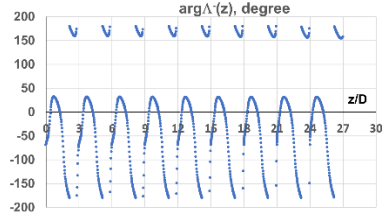

Figure 20 Spatial distributions of the phase of the function $\Lambda^-(z)$.

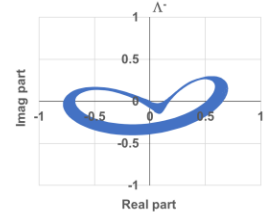

Figure 21 Trajectory in the complex plane of the function $\Lambda^-(z)$.

### 3.2 Beam loading in an inhomogeneous waveguide

Modification of the eigenmodes of a homogeneous waveguide for use in expansion (1) is accomplished by a special choice of the geometric parameters of the base waveguide at each value of longitudinal coordinate $z$ [18,19]. This transforms the piecewise task into the continues one that can be solved using differential equations (2).

In the accelerating section under consideration, four geometric parameters change along the regular part: the thickness of the disks and the length of the resonators, the radii of the holes and resonators [20,23].

When calculating generalized eigenvectors, it is necessary to use generalized geometric parameters, which are selected taking into account the following rules.

To satisfy the boundary conditions, it is necessary to keep the generalized disk thickness and the generalized hole radius constant along each disk, while we can vary the generalized resonator length and the generalized resonator radius. Along each resonator, it is necessary to keep the generalized resonator length and its generalized radius constant, while we can vary the generalized disk thickness and the generalized hole radius (see Figure 22).

Since the derivatives of the generalized geometric parameters $g_i^{(z)}(z)$ vary greatly in the regions of disks and the coupling coefficients are proportional to $dg_i^{(z)}/dz$, $U_{k',k}^{(z)}$ will also vary greatly in these regions (see Figure 23, Figure 24). At the edges of the disk and resonator regions $dg^{(z)}/dz \to 0$, therefore, $U_{k',k}^{(z)}$ also tends to zero. Note that in each region the sum for the coupling coefficients $U_{k',k}^{(z)}$ contains only two terms.

For inhomogeneous waveguide $\tilde{N}_1^{(e,z)} \neq const$, $\gamma_1^{(e,z)} \neq const$ (see Figure 1-Figure 3). As can be seen from (8), that due to the coupling, the generalized longitudinal wave vector $\gamma_1^{(e,z)}$ changes by the value $(-U_{1,1}^{(z)})$. This addition is small, moreover it has an oscillating character (see Figure 23) which reduces its influence on the distribution of phases along the entire section practically to zero. However, the coupling coefficient $U_{\pm1,\pm1}^{(z)}$ s are not small, especially in disk regions (see Figure 24).

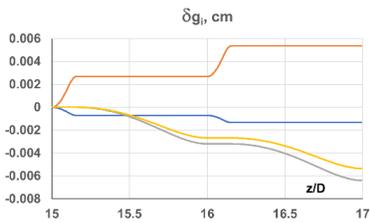

Figure 22 Increment of generalized geometrical parameters along two cells: 1 – generalized cell radius, 2- generalized cell length, 3 - generalized hole radius, 4 – generalized disk thickness.

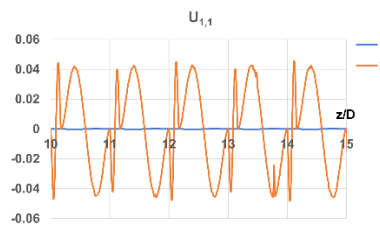

Figure 23 Coupling coefficient $U_{1,1}^{(z)}$ as a function of $z$: 1- $\mathrm{Re}U_{1,1}^{(z)}$, 2- $\mathrm{Im}U_{1,1}^{(z)}$

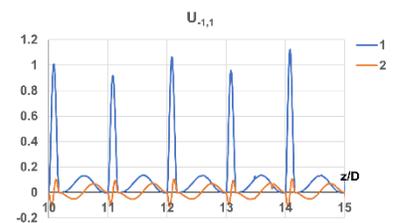

Figure 24 Coupling coefficient $U_{-1,1}^{(z)}$ as a function of $z$: 1- $\mathrm{Re}U_{-1,1}^{(z)}$, 2- $\mathrm{Im}U_{-1,1}^{(z)}$

Results of calculation of the field distribution with using the Runge-Kutta scheme (13) are presented in Figure 25-Figure 31. For greater clarity, we present the calculation results for a homogeneous (denoted as a)) and inhomogeneous (denoted as b)) waveguide simultaneously.

Analysis of the results of calculation of the field distribution for a homogeneous waveguide using the Runge-Kutta scheme (13) and expressions (19),(20) shows that they are in good agreement, taking into account the accuracy of expansion (14) (compare, for example, Figure 21 and Figure 27a)).

From Figure 25b-Figure 26b it follows that the amplitude $C_1(z)$ in an inhomogeneous waveguide behaves similar to the homogeneous case, with the exception of a stronger increase and the occurrence of oscillations.

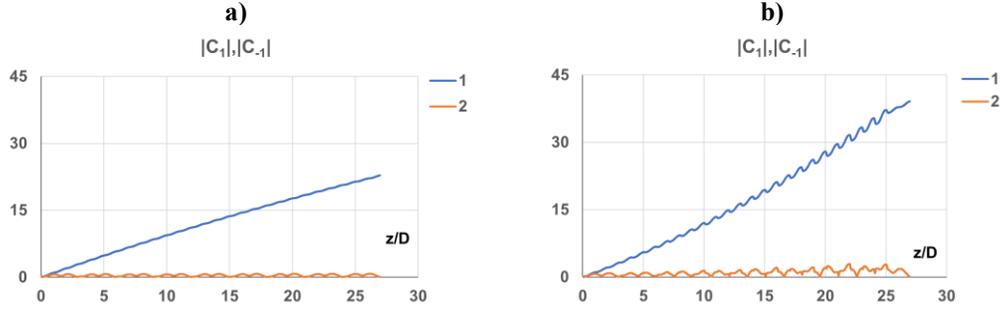

Figure 25 Spatial distributions of the modulus of the amplitudes $C_1$ (1) and $C_{-1}$ (2), $Q$ =0.1 nC, a)- constant impedance case, b) - nonuniform case.

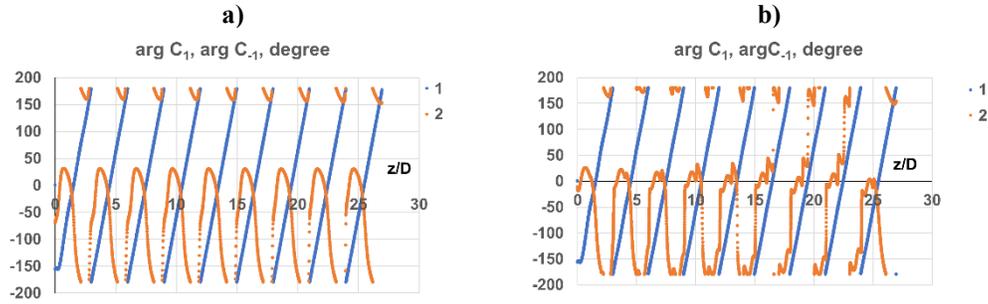

Figure 26 Spatial distributions of the phases of the amplitudes $C_1$ (1) and $C_{-1}$ (2), $Q$ =0.1 nC, a)- constant impedance case, b) - nonuniform case.

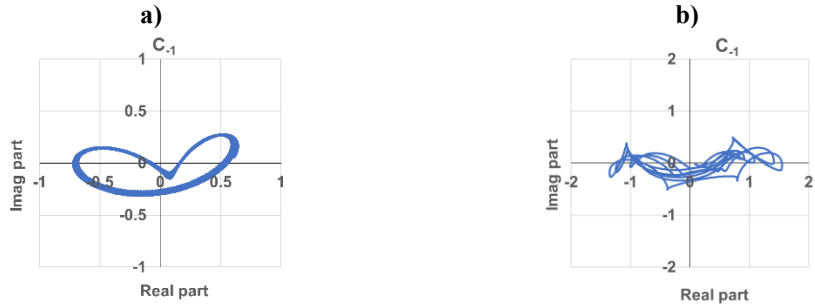

Figure 27 Trajectory in the complex plane of amplitude $C_{-1}(z)$, $Q$ =0.1 nC, a)- constant impedance case, b) - nonuniform case.

As for the amplitude $C_{-1}(z)$, it also has a larger modulus, but the phase distribution has a new feature. When amplitude $C_1(z)$ become large enough, the phase distribution changes – the phase becomes growing function (see Figure 26b, $15 < z/D < 25$). Such phase behavior was first observed in calculations of the field distribution in an inhomogeneous waveguide without a beam [20,21] (for an inhomogeneous medium see [27]). It makes the dependence of $C_{-1}(z)$ on $z$ quite complex (see Figure 27b)

Above we discussed the dependences of amplitudes $C_{\pm1}(z)$ on $z$. To obtain the values of the electric field, we need to multiply these amplitudes by vector functions $\vec{E}_{\pm1}^{(e,z)}(\vec{r})$ (see (5)), which are not periodic for the inhomogeneous waveguide. Trajectories in the complex plane of the components of longitudinal electric field $E_{1,z}^{\pm}(r,z)$ are presented in Figure 28 and Figure 29, the total longitudinal electric field – in Figure 30. They show that despite the complex behavior of component $E_{1,z}^{-}(r,z)$ the total electric field $E_{1,z}(\vec{r}) = E_{1,z}^{+}(\vec{r}) + E_{1,z}^{-}(\vec{r})$ is smooth function along the axis $z$ and has a regular phase shift. This is only possible if the complex behavior of the component $E_{1,z}^{-}(r,z)$ is compensated by the component $E_{1,z}^{+}(r,z)$.

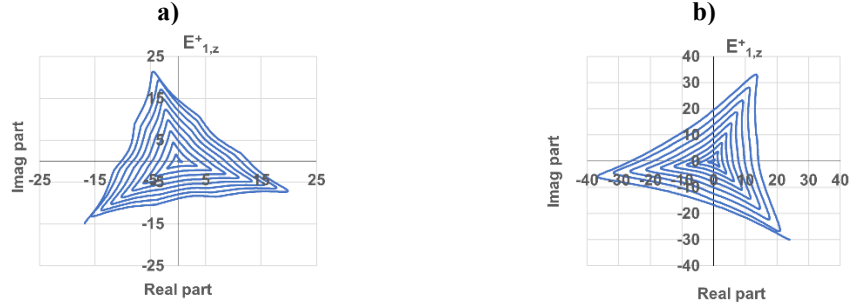

Figure 28 Trajectory in the complex plane of the longitudinal electric field $E_{1,z}^{+}(r,z)\big|_{r=0}$, $Q$ =0.1 nC, a)- constant impedance case, b) - nonuniform case.

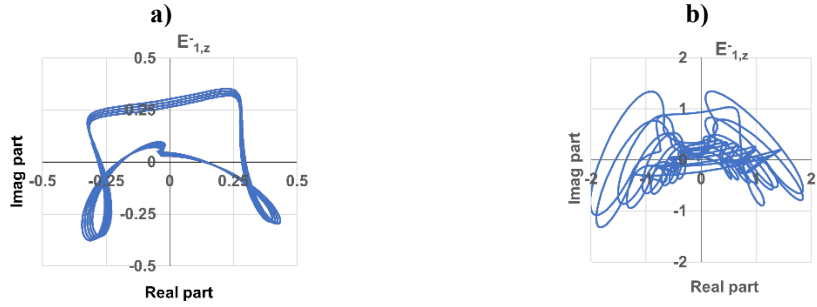

Figure 29 Trajectory in the complex plane of the longitudinal electric field $E_{1,z}^{-}(r,z)\big|_{r=0}$, $Q$ =0.1 nC, a)- constant impedance case, b) - nonuniform case.

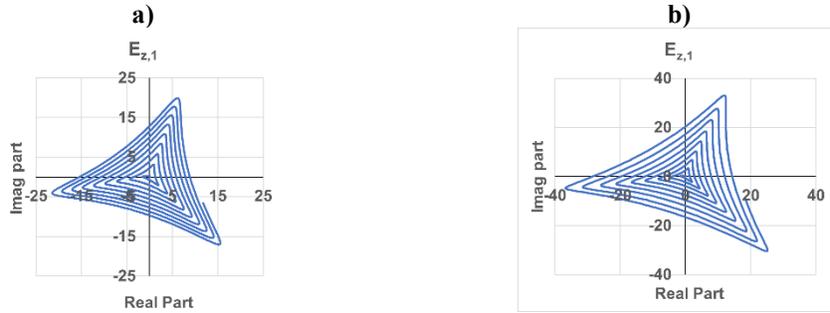

Figure 30 Trajectory in the complex plane of the full longitudinal electric field $E_{1,z}(r,z)\big|_{r=0}$; $Q$ =0.1 nC, a)- constant impedance case, b) - nonuniform case.

The component $E_{1,z}^{-}(r,z)$ constitutes a noticeable part of the total field $E_{1,z}(\vec{r}) = E_{1,z}^{+}(\vec{r}) + E_{1,z}^{-}(\vec{r})$, especially at the beginning of the section. This can be seen from Figure 31. In inhomogeneous waveguide its influence is greater than in homogeneous ones (see also Figure 32). It is connected with the fact that this component is exited not only by electron beam, but by the component $E_{1,z}^{+}(r,z)$ due to their coupling, which is defined by the value of inhomogeneity (terms $U_{\pm1,\mp1}$ in the system (6)). Excitation of the component $E_{1,z}^{-}(r,z)$ is not resonance, but the component $E_{1,z}^{+}(r,z)$ is growing along the axis $z$. This leads to an increase in the component $E_{1,z}^-(r,z)$ (compare Figure 32a and Figure 32b)

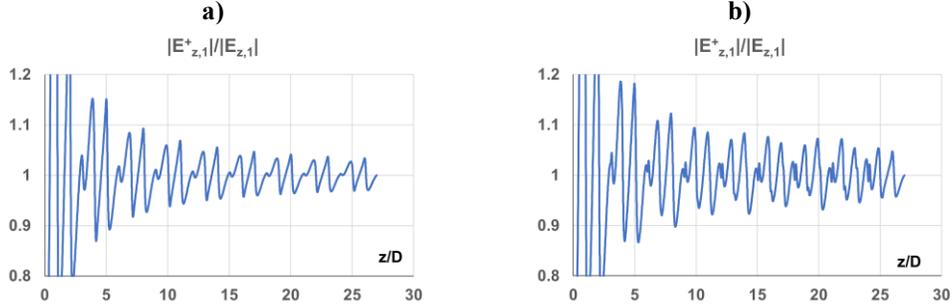

Figure 31 Modulus of deviation in the representation of the longitudinal electric field $E_{z,1}(r=0,z)$ by the field $E_{z,1}^+(r=0,z)$, $Q$ =0.1 nC, a)- constant impedance case, b) - nonuniform case.

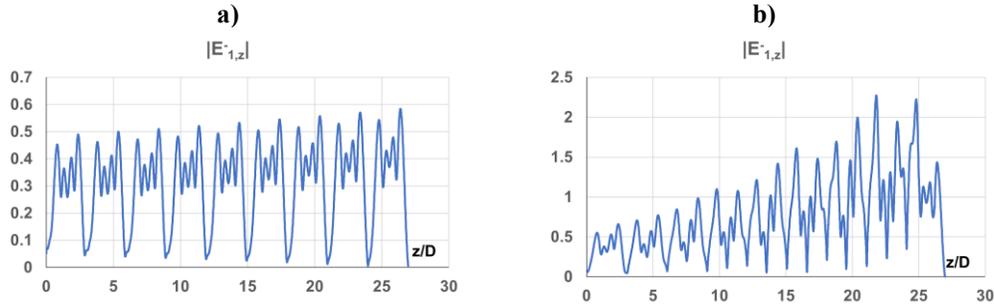

Figure 32 Spatial distributions of the modulus of the longitudinal electric field $E_{1,z}^-(r,z)\big|_{r=0}$, $Q$ =0.1 nC, a)- constant impedance case, b) - nonuniform case.

The question arises about the accuracy of representing the electric field $E_{1,z}(\vec{r})$ only by the component $E_{1,z}^+(\vec{r}) = C_1(z)E_{1,z}^{(e,z)}(\vec{r})$, where $C_1(z)$ is the solution of such an equation

$$\frac{dC_1}{dz} - \gamma_1^{(e,z)}C_1 + \frac{1}{2N_s^{(e,z)}}\frac{dN_s^{(e,z)}}{dz}C_1 = \frac{1}{N_1^{(e,z)}}\int\limits_{S_z^{(z)}}\vec{j}\vec{E}_{-1}^{(e,z)}dS, \qquad (22)$$

which is obtained from the system (6) if we set $U_{s,s'}$ equal zero.

This equation is identical in form to the equations used so far [1-3,7-13], but unlike them all the coefficients $\gamma_1^{(e,z)}(z)$, $N_s^{(e,z)}(z)$, $\vec{E}_{-1}^{(e,z)}(z)$ in this equation are strictly defined as functions of $z$ and are related to the waveguide geometry. It can be shown that for the section under consideration the solution of equation (22) practically coincide with the solution $C_1(z)$ of coupled system (6). It follows from this fact that the accuracy of the representation of the total electric field $E_{1,z}(\vec{r})$ by the component $E_{1,z}^+(\vec{r})$ is determined only by the magnitude of the component $E_{1,z}^-(\vec{r})$. Analysis of Figure 31 shows that the field representation error is not small.

## 4. Conclusions

New generalization of the theory of coupled modes, proposed in [18], gives possibility to construct a self-consistent semi-analytical theory of beam loading in inhomogeneous accelerating structures.

Based on a set of modified eigen functions of a homogeneous periodic waveguide, the total field is represented as a sum of these functions with unknown scalar coefficients that are solutions for an infinitive system of coupled equations. It was reduced and only the single mode approximation was considered when the fields are represented as a sum of two components, one of which is associated with the right travelling eigen wave, and the second with the left. However, this second component is not always a left travelling. When the field is excited by an electron beam it can have complex spatial distribution.

It was shown that for the section under consideration the accuracy of the representation of the total electric field $E_{1,z}(\vec{r})$ by the component $E_{1,z}^+(\vec{r})$ is determined only by the magnitude of the component $E_{1,z}^-(\vec{r})$. Analysis shows that the field representation error is not small.

## APPENDIX 1 Solution of the kinetic equation by the method of characteristics

The simplest kinetic equation describing the time evolution of the distribution function of collisionless beams has the form

$$\frac{\partial f}{\partial t} + \vec{v}\frac{\partial f}{\partial \vec{r}} + \vec{F}\frac{\partial f}{\partial \vec{v}} = 0, \qquad (23)$$

where $f = f(t, \vec{r}, \vec{v})$ is the distribution function and $\vec{F} = \vec{F}(t, \vec{r}, \vec{v}) = q\vec{E} + q[\vec{v}\vec{B}]$. In accelerator physics and physics of electronic devices the boundary task is more interesting when the electron beam is injected into the interaction region and distribution function is known at $z = 0$

$$f(t, z = 0, \vec{r}_\perp, \vec{v}) = f_0(t, \vec{r}_\perp, \vec{v}). \qquad (24)$$

This boundary condition only makes sense if the particles do not stop and always have $v_z > 0$.

In the work [28] the case was considered when all particles in the system have a sufficiently large average velocity and the deviations of the particle velocities from it are small. We will consider the general case.

Let's find the solutions of such system of differential equations

$$\frac{dt}{dz} = \frac{1}{v_z},$$

$$\frac{d\vec{r}_\perp}{dz} = \frac{\vec{v}_\perp}{v_z}, \qquad (25)$$

$$\frac{d\vec{v}}{dz} = \frac{\vec{F}}{v_z}$$

with boundary conditions

$$t(z = 0) = t_0, \vec{r}_\perp(z = 0) = \vec{r}_{\perp,0}), \vec{v}(z = 0) = \vec{v}_0. \qquad (26)$$

We can wright the solution for (25) as

$$t = t_l(t_0, \vec{r}_{\perp,0}, \vec{v}_0, z),$$

$$\vec{r}_\perp = \vec{r}_{\perp,l}(t_0, \vec{r}_{\perp,0}, \vec{v}_0, z), \qquad (27)$$

$$\vec{v} = \vec{v}_l(t_0, \vec{r}_{\perp,0}, \vec{v}_0, z).$$

Inverting this system with respect to the boundary values, we obtain new functions

$$t_0 = t_{l,0}(t, \vec{r}, \vec{v}),$$

$$\vec{r}_{\perp,0} = \vec{r}_{\perp,l,0}(t, \vec{r}, \vec{v}), \qquad (28)$$

$$\vec{v}_0 = \vec{v}_{l,0}(t, \vec{r}, \vec{v}).$$

The functions $t_{l,0}(t, \vec{r}, \vec{v})$, $\vec{r}_{\perp,l,0}(t, \vec{r}, \vec{v})$, $\vec{v}_{l,0}(t, \vec{r}, \vec{v})$ are the first integrals of the system (25) Therefore, the solution to the equation (23) can be written as

$$f = f_0[t_{l,0}(t, \vec{r}, \vec{v}), \vec{r}_{\perp,l,0}(t, \vec{r}, \vec{v}), \vec{v}_{l,0}(t, \vec{r}, \vec{v})], \qquad (29)$$

The system of equations (25) represents the equation of motion of a particle, written relative to the longitudinal coordinate $z$, moving under the action of the force $\vec{F}$.

The function $t_l(t_0, \vec{r}_{\perp,0}, \vec{v}_0, z)$ has a simple physical meaning – it is a moment of time when particle, entering into the considered part of space ($z > 0$) at a moment of time $t = t_0$ with a transverse coordinate $\vec{r}_{\perp,0}$, will arrive at a cross-section with a coordinate $z$

Expression (29) can be written as:

$$f = f_0\left[t_{l,0}(t,\vec{r},\vec{v}), \vec{r}_{\perp,l,0}(t,\vec{r},\vec{v}), \vec{v}_{l,0}(t,\vec{r},\vec{v})\right] =$$

$$\int\limits_{-\infty}^{t} dt_0 \int d\vec{r}_{\perp,0} \int d\vec{v}_0 \, f_0(t_0, \vec{r}_{\perp,0}, \vec{v}_0) \, \delta\left[t_0 - t_{l,0}(t,\vec{r},\vec{v})\right] \times \delta\left[\vec{r}_{\perp,0} - \vec{r}_{\perp,l,0}(t,\vec{r},\vec{v})\right] \times \delta\left[\vec{v}_0 - \vec{v}_{l,0}(t,\vec{r},\vec{v})\right] . \tag{30}$$

The product of delta functions can be rewritten as

$$\delta\left[t_0 - t_{l,0}(t,\vec{r},\vec{v})\right] \times \delta\left[\vec{r}_{\perp,0} - \vec{r}_{\perp,l,0}(t,\vec{r},\vec{v})\right] \times \delta\left[\vec{v}_0 - \vec{v}_{l,0}(t,\vec{r},\vec{v})\right] =$$

$$\left|\frac{D(t_l, \vec{r}_{\perp,l}, \vec{v}_l)}{D(t_0, \vec{r}_{\perp,0}, \vec{v}_0)}\right| \times \delta\left[t - t_l(t_0, \vec{r}_{\perp,0}, \vec{v}_0, z)\right] \times \delta\left[\vec{r}_\perp - \vec{r}_{\perp,l}(t_0, \vec{r}_{\perp,0}, \vec{v}_0, z)\right] \times \delta\left[\vec{v} - \vec{v}_l(t_0, \vec{r}_{\perp,0}, \vec{v}_0, z)\right], \tag{31}$$

where $\dfrac{D(t_l, \vec{r}_{\perp,l}, \vec{v}_l)}{D(t_0, \vec{r}_{\perp,0}, \vec{v}_0)}$ is a Jacobian determinant.

According to Liouville's formula

$$\frac{D(t_l, \vec{r}_{\perp,l}, \vec{v}_l)}{D(t_0, \vec{r}_{\perp,0}, \vec{v}_0)} = C\exp\left[\int\left(\frac{\partial G_t}{\partial t} + \frac{\partial G_x}{\partial x} + \frac{\partial G_y}{\partial y} + \frac{\partial G_{v_x}}{\partial v_x} + \frac{\partial G_{v_y}}{\partial v_y} + \frac{\partial G_{v_z}}{\partial v_z}\right)dz'\right], \tag{32}$$

where

$$G_t = \frac{1}{v_z}, \; G_x = \frac{v_x}{v_z}, \; G_y = \frac{v_y}{v_z}, \; G_{v_x} = \frac{F_x}{v_z}, \; G_{v_y} = \frac{F_y}{v_z}, \; G_{v_z} = \frac{F_z}{v_z}., \tag{33}$$

From these expressions it follows that

$$\frac{\partial G_t}{\partial t} = \frac{\partial G_x}{\partial x} = \frac{\partial G_y}{\partial y} = 0, \tag{34}$$

and

$$\frac{\partial G_t}{\partial t} + \frac{\partial G_x}{\partial x} + \frac{\partial G_y}{\partial y} + \frac{\partial G_{v_x}}{\partial v_x} + \frac{\partial G_{v_y}}{\partial v_y} + \frac{\partial G_{v_z}}{\partial v_z} =$$

$$\frac{1}{v_z}\left[\frac{\partial F_x}{\partial v_x} + \frac{\partial F_y}{\partial v_y}\right] + \frac{\partial}{\partial v_z}\left[\frac{F_z}{v_z}\right] = -\frac{1}{v_z^2}F_z + \frac{1}{v_z}\left[\frac{\partial F_x}{\partial v_x} + \frac{\partial F_y}{\partial v_y} + \frac{\partial F_z}{\partial v_z}\right] = \tag{35}$$

$$= -\frac{1}{v_z}\frac{\partial v_z}{\partial z} + \frac{1}{v_z}\left[\frac{\partial F_x}{\partial v_x} + \frac{\partial F_y}{\partial v_y} + \frac{\partial F_z}{\partial v_z}\right] = -\frac{\partial \ln(v_z)}{\partial z} + \frac{1}{v_z}\left[\frac{\partial F_x}{\partial v_x} + \frac{\partial F_y}{\partial v_y} + \frac{\partial F_z}{\partial v_z}\right]$$

because according to (25) $\dfrac{F_z}{v_z} = \dfrac{\partial v_z}{\partial z}$.

The force $\vec{F}$ has such form

$$\vec{F} = q\vec{E} + q\left[\vec{v}\times\vec{B}\right] = q\vec{E} + q\begin{bmatrix} \vec{e}_x & \vec{e}_y & \vec{e}_z \\ v_x & v_y & v_z \\ B_x & B_y & B_z \end{bmatrix}, \tag{36}$$

from which it follows that

$$\frac{\partial F_x}{\partial v_x} + \frac{\partial F_y}{\partial v_y} + \frac{\partial F_z}{\partial v_z} = 0 \tag{37}$$

and

$$\frac{D(t_l, \vec{r}_{\perp,l}, \vec{v}_l)}{D(t_0, \vec{r}_{\perp,0}, \vec{v}_0)} = C\exp\left[-\int\limits_0^z dz' \frac{\partial \ln(v_z)}{\partial z'}\right] = C\frac{v_{z,0}}{v_z}. \tag{38}$$

Since at $z = 0$ the boundary conditions (26) are satisfied, then

$$C = \frac{D(t_l, \vec{r}_{\perp,l}, \vec{v}_l)}{D(t_0, \vec{r}_{\perp,0}, \vec{v}_0)}\bigg|_{z=0} = 1. \tag{39}$$

Finally, we get

$$f = \int\limits_{-\infty}^{t} d\,t_0 \int d\,\vec{r}_{\perp,0} \int d\,\vec{v}_0\, f_0(t_0, \vec{r}_{\perp,0}, \vec{v}_0) \frac{v_{z,0}}{v_z}$$
$$\delta\left[t - t_l(t_0, \vec{r}_{\perp,0}, \vec{v}_0, z)\right] \times \delta\left[\vec{r}_\perp - \vec{r}_{\perp,l}(t_0, \vec{r}_{\perp,0}, \vec{v}_0, z)\right] \times \delta\left[\vec{v} - \vec{v}_l(t_0, \vec{r}_{\perp,0}, \vec{v}_0, z)\right]$$
(40)

If we assume that $\vec{r}_{\perp,l}(t_0, \vec{r}_{\perp,0}, \vec{v}_0, z) = \vec{r}_{\perp,0}$, $f_0(t_0, \vec{r}_{\perp,0}, \vec{v}_0) = f_0(t_0, \vec{r}_{\perp,0})$, $|v_z - v_{z,0}| \le v_{z,0}$ and accept that $v_z = v_{z,0}$, we will get the results of the work [28]. Another method of solving the kinetic equation (23) is proposed in [29]

The charge and current densities are expressed through the distribution function as follows

$$\rho = \int f\, d\,\vec{v} = \int d\,\vec{v} \int\limits_{-\infty}^{t} d\,t_0 \int d\,\vec{r}_{\perp,0} \int d\,\vec{v}_0\, f_0(t_0, \vec{r}_{\perp,0}, \vec{v}_0) \frac{v_{z,0}}{v_z} \times$$
$$\delta\left[t - t_l(t_0, \vec{r}_{\perp,0}, \vec{v}_0, z)\right] \delta\left[\vec{r}_\perp - \vec{r}_{\perp,l}(t_0, \vec{r}_{\perp,0}, \vec{v}_0, z)\right] \delta\left[\vec{v} - \vec{v}_l(t_0, \vec{r}_{\perp,0}, \vec{v}_0, z)\right] =$$
$$\int\limits_{-\infty}^{t} d\,t_0 \int d\,\vec{r}_{\perp,0} \int d\,\vec{v}_0\, f_0(t_0, \vec{r}_{\perp,0}, \vec{v}_0) \frac{v_{z,0}}{v_{l,z}(t_0, \vec{r}_{\perp,0}, \vec{v}_0, z)} \delta\left[t - t_l(t_0, \vec{r}_{\perp,0}, \vec{v}_0, z)\right] \delta\left[\vec{r}_\perp - \vec{r}_{\perp,l}(t_0, \vec{r}_{\perp,0}, \vec{v}_0, z)\right]$$
(41)

$$\vec{j} = \int \vec{v} f\, d\,\vec{v} = \int d\,\vec{v} \int\limits_{-\infty}^{t} d\,t_0 \int d\,\vec{r}_{\perp,0} \int d\,\vec{v}_0\, f_0(t_0, \vec{r}_{\perp,0}, \vec{v}_0) \frac{v_{z,0}}{v_z} \vec{v} \times$$
$$\delta\left[t - t_l(t_0, \vec{r}_{\perp,0}, \vec{v}_0, z)\right] \delta\left[\vec{r}_\perp - \vec{r}_{\perp,l}(t_0, \vec{r}_{\perp,0}, \vec{v}_0, z)\right] \delta\left[\vec{v} - \vec{v}_l(t_0, \vec{r}_{\perp,0}, \vec{v}_0, z)\right] =$$
$$\int\limits_{-\infty}^{t} d\,t_0 \int d\,\vec{r}_{\perp,0} \int d\,\vec{v}_0\, f_0(t_0, \vec{r}_{\perp,0}, \vec{v}_0) \frac{v_{z,0}}{v_{l,z}(t_0, \vec{r}_{\perp,0}, \vec{v}_0, z)} \vec{v}_l(t_0, \vec{r}_{\perp,0}, \vec{v}_0, z) \delta\left[t - t_l(t_0, \vec{r}_{\perp,0}, \vec{v}_0, z)\right] \delta\left[\vec{r}_\perp - \vec{r}_{\perp,l}(t_0, \vec{r}_{\perp,0}, \vec{v}_0, z)\right]$$
(42)

According to definition (41) the dimension of $f$ - $[f] = qT^3 / L^6$ .

At the input ( $z = 0$ ) the current is equal to

$$\vec{j}_0 = \int\limits_{-\infty}^{t} d\,t_0 \int d\,\vec{r}_{\perp,0} \int d\,\vec{v}_0\, f_0(t_0, \vec{r}_{\perp,0}, \vec{v}_0) \vec{v}_0\, \delta\left[t - t_0\right] \delta\left[\vec{r}_\perp - \vec{r}_{\perp,0}\right] = \int f_0(t, \vec{r}_\perp, \vec{v}_0) \vec{v}_0\, d\,\vec{v}_0 \quad.$$
(43)

For a sequence of point monoenergetic bunches with $v = \vec{e}_z V_0$

$$f_0(t, \vec{r}_\perp, \vec{v}_0) = \delta(x)\delta(y)\delta(v_{0,z} - V_0)\delta(v_{0,x})\delta(v_{0,y}) \frac{1}{V_0} \sum_n Q_n \delta\left(t - nT_0\right)$$
(44)

and

$$j_z = \int v_z f\, d\,\vec{v} = \delta(x)\delta(y)\sum_n Q_n \delta\left(t - \frac{z}{V_0} - nT_0\right).$$
(45)

Fourier component of this current has the form

$$j_{z,\omega} = \int\limits_{-\infty}^{\infty} j_z \exp\left(i\omega t\right) dt = \delta(x)\delta(y)\exp\left\{i\omega \frac{z}{V_0}\right\} \sum_n Q_n \exp\left(i\omega n T_0\right).$$
(46)

If $Q_n$ is constant $Q_n = Q$ then since $\sum\limits_{n=-\infty}^{\infty} \exp\left(i\omega n T_0\right) = \frac{1}{T_0} \sum\limits_{n=-\infty}^{\infty} \delta\left(\omega - n\omega_0\right)$ we get

$$j_{z,\omega} = \delta(x)\delta(y)\exp\left\{i\omega \frac{z}{V_0}\right\} \frac{Q}{T_0} \sum\limits_{n=-\infty}^{\infty} \delta\left(\omega - n\omega_0\right)$$
(47)

## APPENDIX 2 Runge-Kutta method

Here we will give the formulas that we used to solve the coupled equations (9), which we rewrite as

$$\frac{dC_1}{dz} = f_1(z)C_2 + F_1(z),$$
$$\frac{dC_2}{dz} = f_2(z)C_1 + F_2(z).$$
(48)

If we know the function values $C_{1,n}, C_{2,n}$ at $z = z_n$, then the function values $C_{1,n+1}, C_{2,n+1}$ at $z = z_n + h$ can be found by using the Runge-Kutta formulas

$$C_{1,n+1} = \alpha_{1,1}(n)G_{1,n} + \alpha_{1,2}(n)G_{2,n} + W_{1,n},$$
$$C_{2,n+1} = \alpha_{2,1}(n)G_{1,n} + \alpha_{2,2}(n)G_{2,n} + W_{2,n},$$

(49)

where

$$\alpha_{1,1}(n) = 1 + \frac{1}{6}\left(k_{0,1}^{(1)} + 2k_{1,1}^{(1)} + 2k_{2,1}^{(1)} + k_{3,1}^{(1)}\right)$$

$$\alpha_{1,2}(n) = \frac{1}{6}\left(k_{0,1}^{(2)} + 2k_{1,1}^{(2)} + 2k_{2,1}^{(2)} + k_{3,1}^{(2)}\right)$$

$$\alpha_{2,1}(n) = \frac{1}{6}\left(k_{0,2}^{(1)} + 2k_{1,2}^{(1)} + 2k_{2,2}^{(1)} + k_{3,2}^{(1)}\right)$$

(50)

$$\alpha_{2,2}(n) = 1 + \frac{1}{6}\left(k_{0,2}^{(2)} + 2k_{1,2}^{(2)} + 2k_{2,2}^{(2)} + k_{3,2}^{(2)}\right)$$

$$W_{1,n} = \frac{1}{6}\left(K_{0,1} + 2K_{1,1} + 2K_{2,1} + K_{3,1}\right)$$

$$W_{2,n} = \frac{1}{6}\left(K_{0,2} + 2K_{1,2} + 2K_{2,2} + K_{3,2}\right)$$

(51)

$$k_{0,1}^{(1)} = 0; \quad k_{0,1}^{(2)} = hf_1(z_n)$$

$$k_{1,1}^{(1)} = hf_1\left(z_n + \frac{h}{2}\right)\frac{k_{0,2}^{(1)}}{2}; \quad k_{1,1}^{(2)} = hf_1\left(z_n + \frac{h}{2}\right)$$

$$k_{2,1}^{(1)} = hf_1\left(z_n + \frac{h}{2}\right)\frac{k_{1,2}^{(1)}}{2}; \quad k_{2,1}^{(2)} = hf_1\left(z_n + \frac{h}{2}\right)\left(1 + \frac{k_{1,2}^{(2)}}{2}\right)$$

(52)

$$k_{3,1}^{(1)} = hf_1(z_n + h)k_{2,2}^{(1)}; \quad k_{3,1}^{(2)} = hf_1(z_n + h)\left(1 + k_{2,2}^{(2)}\right)$$

$$k_{0,2}^{(1)} = hf_2(z_n); \quad k_{0,2}^{(2)} = 0$$

$$k_{1,2}^{(1)} = hf_2\left(z_n + \frac{h}{2}\right); \quad k_{1,2}^{(2)} = hf_2\left(z_n + \frac{h}{2}\right)\frac{k_{0,1}^{(2)}}{2}$$

$$k_{2,2}^{(1)} = hf_1\left(z_n + \frac{h}{2}\right)\left(1 + \frac{k_{1,1}^{(2)}}{2}\right); \quad k_{2,2}^{(2)} = hf_1\left(z_n + \frac{h}{2}\right)\frac{k_{1,1}^{(2)}}{2}$$

(53)

$$k_{3,2}^{(1)} = hf_1(z_n + h)\left(1 + k_{2,1}^{(1)}\right); \quad k_{3,2}^{(2)} = hf_1(z_n + h)k_{2,1}^{(2)}$$

$$K_{0,1} = hF_1(z_n)$$

$$K_{1,1} = hf_1\left(z_n + \frac{h}{2}\right)\frac{1}{2}K_{0,2} + hF_1\left(z_n + \frac{h}{2}\right);$$

$$K_{2,1} = hf_1\left(z_n + \frac{h}{2}\right)\frac{1}{2}K_{1,2} + hF_1\left(z_n + \frac{h}{2}\right);$$

$$K_{3,1} = hf_1(z_n + h)K_{2,2} + hF_1(z_n + h);$$

$$K_{0,2} = hF_2(z_n)$$

(54)

$$K_{1,2} = hf_2\left(z_n + \frac{h}{2}\right)\frac{1}{2}K_{0,1} + hF_2\left(z_n + \frac{h}{2}\right);$$

$$K_{2,2} = hf_2\left(z_n + \frac{h}{2}\right)\frac{1}{2}K_{1,1} + hF_2\left(z_n + \frac{h}{2}\right);$$

$$K_{3,2} = hf_2(z_n + h)K_{2,1} + hF_2(z_n + h);$$